**Controlling the phase locking of unstable magnetic bits for ultra-low power computation**

A. Mizrahi[1,2,*], N. Locatelli[2], R. Lebrun[1], V. Cros[1], A. Fukushima[3], H. Kubota[3], S. Yuasa[3], D. Querlioz[2] and J. Grollier[1]

1 - Unité Mixte de Physique CNRS, Thales, Univ. Paris-Sud, Université Paris-Saclay, 91767 Palaiseau, France

2 - Institut d'Electronique Fondamentale, Univ. Paris-Sud, CNRS, Université Paris-Saclay, 91405 Orsay, France

3 - Spintronics Research Center, National Institute of Advanced Industrial Science and Technology (AIST), Tsukuba, Japan

* alicecmmizrahi@gmail.com

**When fabricating magnetic memories, one of the main challenges is to maintain the bit stability while downscaling. Indeed, for magnetic volumes of a few thousand $nm^3$, the energy barrier between magnetic configurations becomes comparable to the thermal energy at room temperature. Then, switches of the magnetization spontaneously occur. These volatile, superparamagnetic nanomagnets are generally considered useless. But what if we could use them as low power computational building blocks? Remarkably, they can oscillate without the need of any external dc drive, and despite their stochastic nature, they can beat in unison with an external periodic signal. Here we show that the phase locking of superparamagnetic tunnel junctions can be induced and suppressed by electrical noise injection. We develop a comprehensive model giving the conditions for synchronization, and predict that it can be achieved with a total energy cost lower than $10^{-13}$ J. Our results open the path to ultra-low power computation based on the controlled synchronization of oscillators.**

**Introduction**

Superparamagnetic tunnel junctions present a number of advantages for computation. First, they can be downscaled to atomic dimensions[1]. In addition, because the energy barrier separating the two magnetic configurations is small, low current densities can lead to significant action of spin torques[2]. But how can they be harnessed for applications? A first option is to use superparamagnetic tunnel junctions as sensors. Indeed, thanks to their high sensitivity to electrical currents they are able to detect weak oscillating signals[3–5] through the effect of stochastic resonance[6]. A second option is to use them as building blocks of computing systems leveraging the synchronization of oscillators for processing[7,8]. It has been recently recognized that coupled nano-oscillators are promising brain-inspired systems for performing cognitive tasks such as pattern recognition[9–16]. Like neurons in some parts of the brain, they compute by synchronizing and desynchronizing depending on sensory inputs[17]. However, such systems require a high number of oscillators, each powered by substantial dc current. Using superparamagnetic tunnel junctions would allow orders of magnitude gain in power consumption. In addition, by shrinking their dimensions they can be fabricated from the same magnetic stack as stable junctions, allowing for densely interweaving oscillations and memory.

Nevertheless there are a number of prerequisites to be able to use superparamagnetic tunnel junctions for computational purposes. In particular, it is necessary to identify handles providing control over their synchronization and to model accurately the associated physics for simulating large scale systems of interacting oscillators. Here we show experimentally that we can induce the phase-locking of a superparamagnetic tunnel junction to a weak periodic signal through the addition of a small electrical noise, and that we can suppress the phase-locking by adding more noise. While the stochastic behavior of most systems becomes unpredictable when shrunk to nanometer scale, the dedicated model we develop here encompasses all our experimental results. The quantitative agreement between model and experiments allows predicting the power consumption of computing systems harnessing phase-locking of superparamagnetic tunnel junctions.

**Experimental Results**

We study experimentally superparamagnetic tunnel junctions with an MgO barrier and a CoFeB free layer of dimensions 60×120×1.7 nm$^3$ (details in Methods). As depicted in the inset of Fig. 1a, we evaluate their ability to phase lock to a weak square periodic drive voltage in the presence of electrical white noise, at room temperature. We set the drive frequency at $F_{ac}$ = 50 Hz and the drive amplitude at $V_{ac}$ = 63 mV, which corresponds to approximately 25% of the voltage threshold for deterministic magnetization switching at 0 K. Fig. 1a shows how the mean frequency of the stochastic oscillator evolves when the amplitude of the electrical noise is increased.

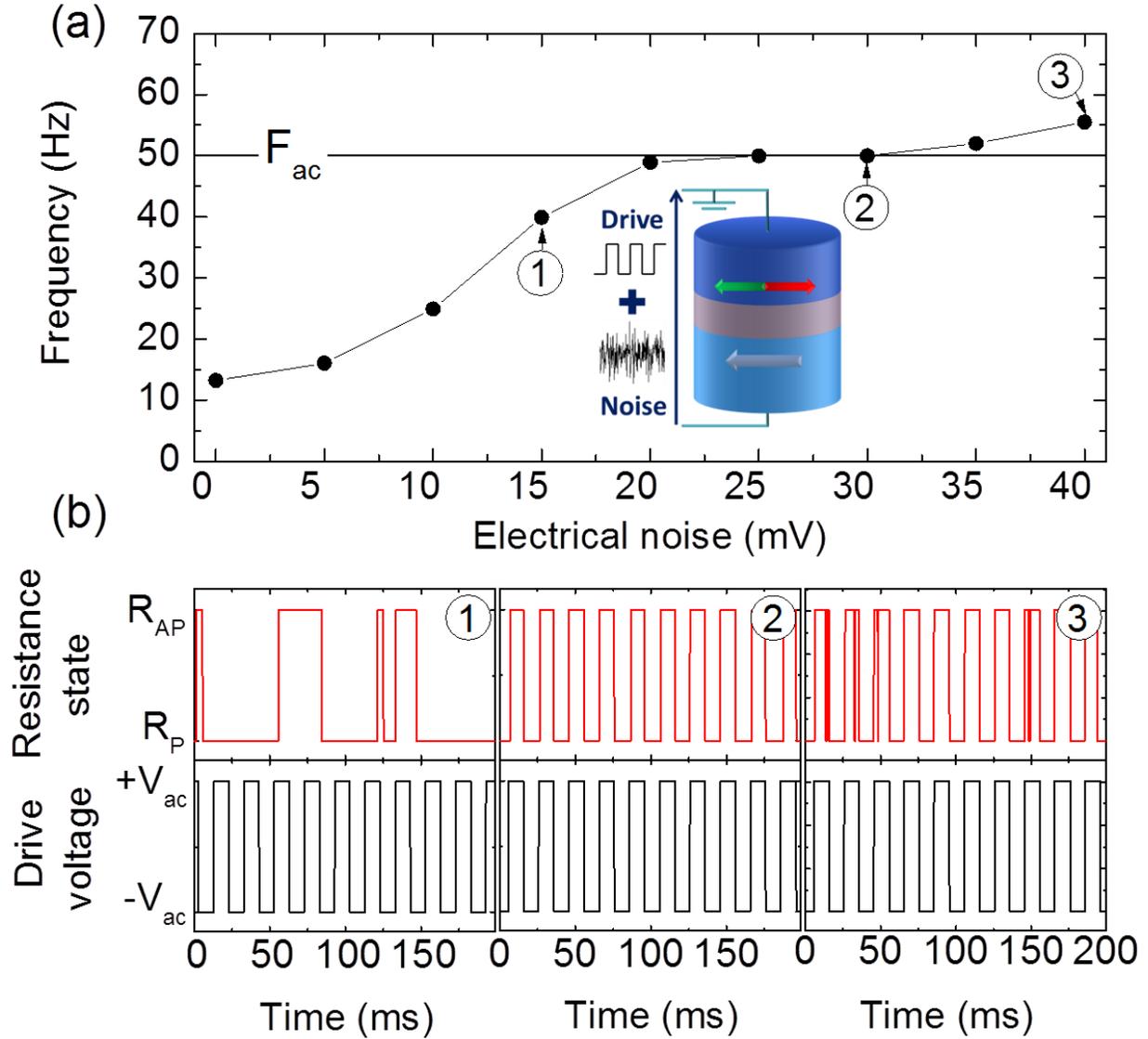

**Fig. 1: Controlling the phase locking of a superparamagnetic tunnel junction through electrical noise: experimental results.** A *square periodic voltage of amplitude $V_{ac} = 63\ mV$ and frequency $F_{ac} = 50\ Hz$ as well as white Gaussian electrical noise are applied to the junction. (a) Inset: schematic of the superparamagnetic tunnel junction driven by a periodic square voltage and electrical noise. Main: junction's mean frequency as a function of electrical noise amplitude (standard deviation $\sigma_{Noise}$). (b) Times traces of the junction's resistance (top) and applied voltage (bottom) for three different levels of noise with standard deviations: (1) $\sigma_{Noise} = 15\ mV$, (2) $\sigma_{Noise} = 30\ mV$ and (3) $\sigma_{Noise} = 40\ mV$.*

We observe three different regimes, illustrated in Fig. 1b. As can be seen in the first panel, the jumps in the junction resistance, corresponding to reversals of the magnetization, remain stochastic for small values of injected electrical noise. In addition, the junction mean frequency is lower than the drive frequency. Usually, adding noise to a system tends to destroy its coherence and is detrimental to the occurrence of a synchronized regime. On the contrary, in our case, by increasing the electrical noise amplitude, we can increase the junction's mean frequency towards the drive frequency. Eventually, for an optimal range of electrical noise (between 20 and 30 mV), we observe both frequency locking (as evidenced from the plateau in Fig. 1a), and phase locking to the driving signal (as shown in panel 2). In this second regime, electrical noise optimally assists the periodic drive to overcome the voltage threshold for magnetization switching at every oscillation of the drive voltage[18–20]. In the third regime (panel 3), higher amplitude electrical noise induces unwanted switches of the magnetization and prevents synchronization.

**Analytical model and simulations of phase-locking**

Noise-controlled phase-locking has been experimentally demonstrated in a few non-linear systems such as Schmitt triggers[19,21] or lasers[22] but never in a nanoscale system as achieved here. In order to assess the potential of superparamagnetic tunnel nanojunctions for applications, we now propose a model that accurately describes their noise-mediated synchronization to weak periodic signals. The thermally activated escape rate of a single domain magnetization, modulated by spin transfer torque[23,24], has a simple expression[25–27]:

$$\phi_{AP \to P \ (P \to AP)} = \phi_0 exp\left(-\frac{\Delta E}{k_B T}\left(1 \pm \frac{V(t)+V_N(t)}{V_c}\right)\right) (1),$$

where $\phi_0$ is the attempt frequency, $\Delta E$ the energy barrier between the two stable states, $T$ the temperature and $V_c$ the voltage threshold for deterministic switching [26,27]. In our case the driving force is the sum of the periodic voltage $V(t) = \pm V_{ac}$ and the electrical noise $V_N(t)$, which is assumed Gaussian with standard deviation $\sigma_{Noise}$. In consequence there are two sources of noise in our system: electrical and thermal noises (T=300K). Using Eq. (1), we can numerically compute the junction's mean frequency as a function of the electrical noise amplitude[28] (see Methods). Fig.2a compares the experimental data (symbols) measured for different amplitudes of the periodic drive to the results of numerical simulations (solid lines). All simulations have been performed using a single set of fitting parameters ($V_c$ = 235mV and $\Delta E/k_B T$ = 22.5), emphasizing the remarkable agreement with experimental results.

The analytical models that have been developed in the past to describe noise-induced phase locking [18,20,29] focused on cases for which noise can be taken into account as a time-independent variable in the escape rates, such as temperature in Eq. (1). However in our case, the escape rates from the parallel and antiparallel states are time varying, random variables because they depend on the electrical noise $V_N(t)$. In order to go further, we develop an original and generic method to analytically determine the conditions for synchronization. Starting from Eq. (1), we calculate the probabilities $P_+$ ($P_-$) for the magnetization to switch from out-of-phase with the drive voltage to in-phase (from in-phase to out-of-phase) during half a period. The details of the derivation and the expressions for the phase-locking and phase-unlocking probabilities $P_+$ and $P_-$ are given in Methods. In the vicinity of the plateau, the mean frequency of the junction is described by $F = F_{ac}(2P_+ + 2P_- - 1)$. Considering a 99% frequency locking requirement, the boundaries of the synchronization region are given by $P_+ > 99.5\%$ and $P_- < 0.5\%$, as shown by the red arrows in Fig.2a for the case of a drive amplitude of 63 mV. Fig. 2b and c show that our analytical model (dotted lines) quantitatively predicts the boundaries of the experimental synchronization zone (symbols) over the whole range of investigated parameters. As can be seen in Fig. 2b, the range of electrical noise for which phase-locking occurs increases with the drive amplitude. When the drive amplitude is too low (here below 37 mV), synchronization cannot be achieved. On the other hand, at large drive amplitudes (here above 85 mV), phase locking can be achieved through room temperature thermal noise alone, without the need to add up electrical noise. In addition, as shown in Fig. 2c, phase-locking is achievable for frequencies orders of magnitude higher than the natural mean frequency (0.1 Hz here).

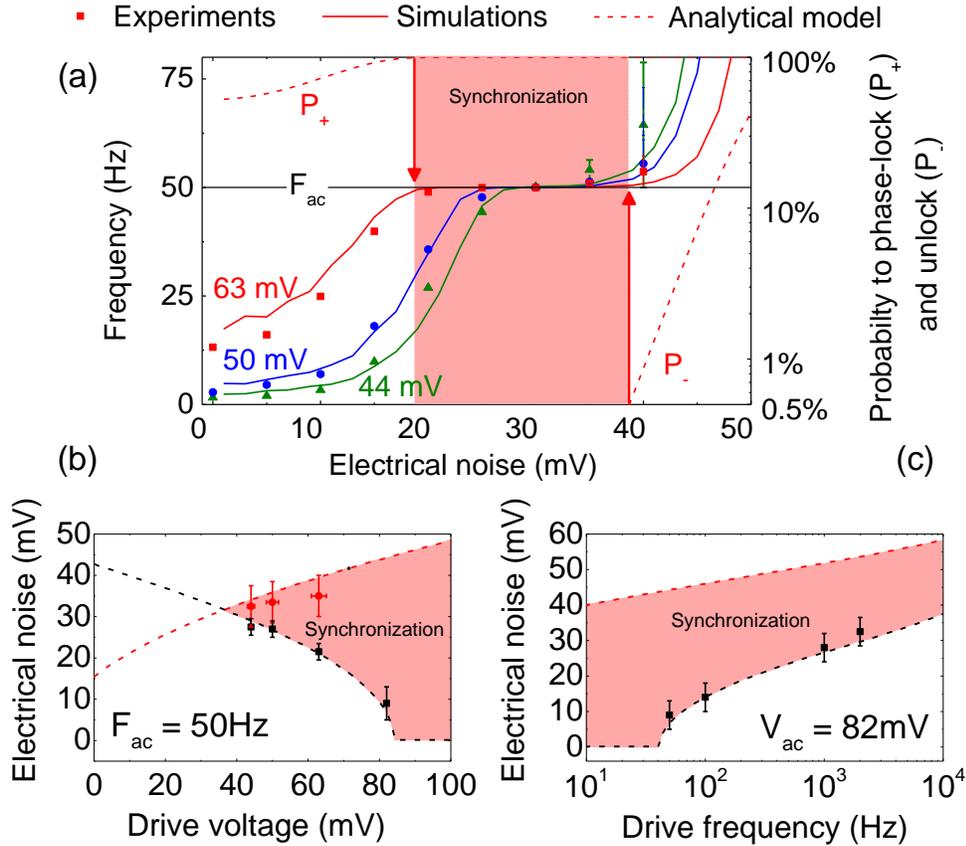

**Fig. 2: Modelling the phase locking of superparamagnetic tunnel junctions to an external periodic drive in the presence of electrical noise.** *Simulations and analytical calculations are done with the same set of parameters: $V_c$ = 235mV and $\Delta E/k_B T$ = 22.5. (a) A square periodic voltage of frequency $F_{ac} = 50\ Hz$ and a white Gaussian electrical noise are applied to a magnetic tunnel junction. Three amplitudes are studied: $V_{ac} = 44\ mV$ (green), $V_{ac} = 50\ mV$ (blue) and $V_{ac} = 63\ mV$ (red). Left axis: frequency of the oscillator versus the standard deviation of the noise, both experimental results (circles, squares and triangles) and numerical results (solid lines) are represented. Right axis: analytical values of probabilities $P_+$ and $P_-$ to switch during half a period $T_{ac}/2$ versus noise (dash lines). Vertical dot lines represent the noise levels for which $P_+ = 99.5\%$ and $P_- = 0.5\%$ for a $63\ mV$ amplitude. The horizontal black solid line represents the drive frequency $F_{ac}$. (b-c) Lower noise bound (black) and higher noise bound (red) of the synchronization plateau versus the drive voltage (b) and versus the drive frequency (c). Both analytical values (dash lines) and experimental results (circles and squares) are presented. In the red zones the oscillator is synchronized with the excitation.*

**Estimation of the energy needed to synchronize superparamagnetic tunnel junctions**

Having validated our analytical model, we can now predict the energy consumption of spintronic circuits leveraging the synchronization of superparamagnetic tunnel junctions for computing. In such circuits, a calculation is finished once steady synchronization patterns are formed within the assembly of oscillators after its perturbation by an external input signal[15,16]. Superparamagnetic tunnel junctions can phase lock fast, typically in a single period of the input signal[30]. To evaluate the energy needed for such operation, we focus on the most recent generation of magnetic tunnel junctions with perpendicularly magnetized layers. We consider junctions small enough (< 30 nm) for their free layer to behave as a macrospin[2] and to be described by Eq. (1). Using parameters (energy barrier, critical voltage and resistance) determined from experiments by Sato et al[2], we calculate the minimum energy $E_{min}$ necessary to synchronize the junction with and without the help of electrical noise.

Figure 3 shows the evolution of $E_{min}$ as a function of junction diameter for different drive frequencies. When only thermal noise is used, $E_{min} = \left(\frac{V_1^2}{R}\right) T_{ac}$ where $V_1$ is the minimum drive voltage required to phase-lock the junction in one drive period $T_{ac}$ (see inset in Fig. 3 and Methods). When electrical noise is added, the minimum energy becomes $E_{min} = \left(\frac{V_0^2}{R} + \frac{\sigma_0^2}{R}\right) T_{ac}$, where $V_0$ is the minimum drive voltage required for phase-locking and $\sigma_0$ the corresponding noise level (see inset in Fig. 3 and Methods). Interestingly, for each drive frequency, there is an optimal diameter $D_{min}$ for which the energy needed to achieve phase-locking is minimal. Indeed the junction diameter determines its natural frequency: large diameters correspond to low frequencies because large magnetic volumes are more difficult to switch. Above $D_{min}$, the drive frequency is larger than the junction's mean frequency. To phase-lock, the junction has to be accelerated. In the absence of electrical noise, this can be done through an increase of the drive amplitude, which enhances the ability of the junction to synchronize (stars in Fig.3).

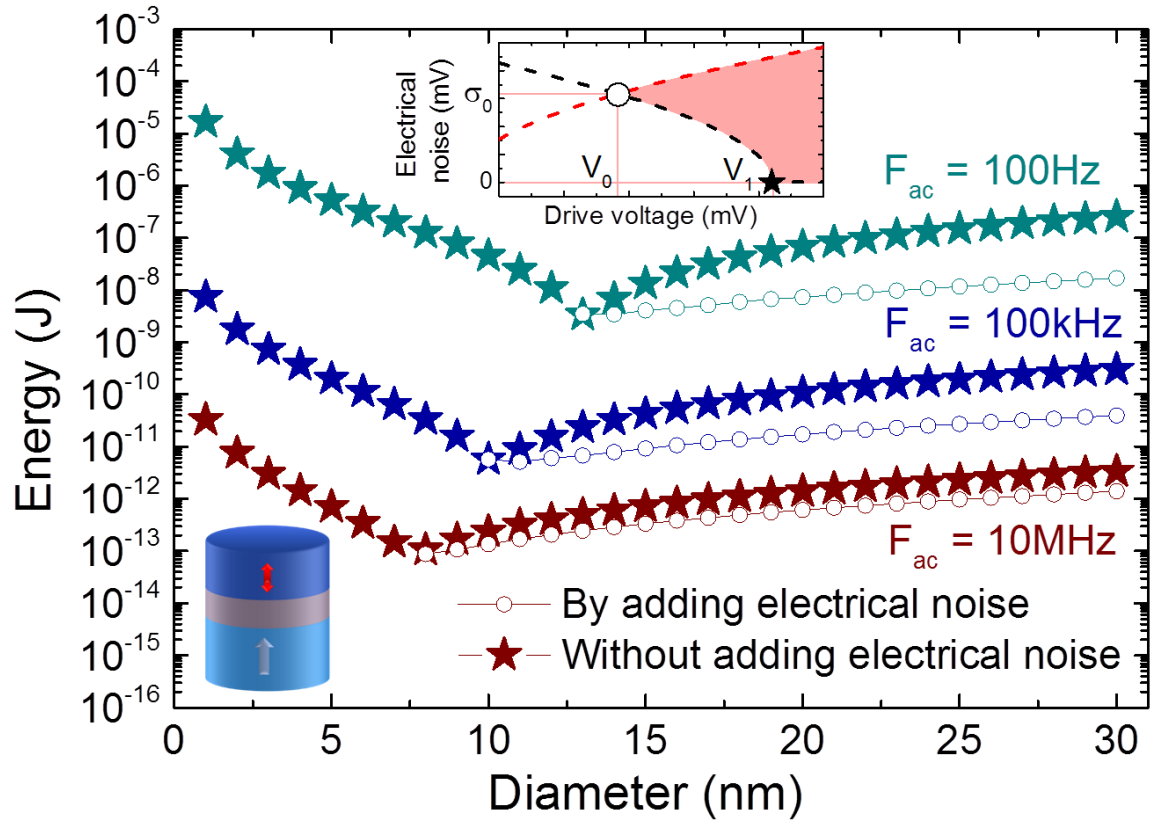

**Figure 3: Energy required to phase-lock a perpendicularly magnetized superparamagnetic tunnel junction: predictions of the analytical model.** *Upper inset: Reproduction of Figure 2b. The circle indicates the lowest drive voltage $V_0$ for which synchronization can be achieved and the corresponding electrical noise level $\sigma_0$. The star indicates the lowest drive voltage $V_1$ for which synchronization can be achieved through thermal noise alone without addition of any electrical noise. Lower inset: schematic view of a perpendicularly magnetized tunnel junction. Main: Calculated minimum energy required to synchronize a perpendicularly magnetized superparamagnetic tunnel junction to a periodic voltage drive in one period, plotted versus the diameter of the junction, for different drive frequencies. Circles represent the case where electrical noise has been added while stars represent the case where only thermal noise is used.*

Adding electrical noise lowers the drive amplitude required to synchronize and thus decreases the total amount of energy to provide (circles in Fig. 3). Below $D_{min}$, the junction has to be slowed down in order to phase lock. As electrical noise always speeds up the oscillator by increasing the number of switches, this can only be achieved by increasing the drive amplitude. Our results indicate that carefully engineering the junctions' dimensions can drastically decrease the energy required to achieve phase-locking, about $8\times10^{-14}$ J for a drive frequency of 10 MHz (Figure 3). By comparison, synchronizing a harmonic dc-driven spin-torque oscillator with a 10 GHz frequency[31] to a drive current would require 100 times more energy (see Methods). CMOS implementations of oscillators for bio-inspired computing applications are also more costly in terms of energy, with a consumption above $7\times10^{-12}$ J for integrate and fire neurons[32]. In addition they occupy a large area on chip, typically several hundreds of µm².

Because of these issues of size and energy consumptions, bio-inspired computing systems leveraging the synchronization of coupled oscillators for computing have never been implemented in CMOS. Thanks to their small area and low energy consumption, arrays of phase-locked superparamagnetic tunnel junctions are a promising alternative for pattern recognition. We take the example of image classification, which generally requires one oscillator per pixel[15,16] to evaluate the energy consumption of magnetic processor leveraging the synchronization of superparamagnetic tunnel junctions. Using the figures determined above, we predict that a superparamagnetic spintronic circuit can classify 1Mpixel images while consuming less than 0.1 µJ. Our results open the way to ultra-low power stochastic computation harnessing superparamagnetism.

**Methods:**

**Sample:** The samples are in-plane magnetized magnetic tunnel junctions. They were fabricated by sputtering, with the stack: substrate ($SiO_2$)/ buffer layer 35 nm / IrMn 7 nm / CoFe 2.5 nm / Ru 0.85 nm / CoFeB 2.4 nm / MgO-barrier 1.0 nm / CoFeB 1.7 nm / capping layer 14 nm. The whole stack was annealed before microfabrication at 300°C under a magnetic field of 1 Tesla for 1 hour. Patterning was then performed by e-beam lithography, resulting in nanopillars with elliptic 60 x 120 nm² cross-sections.

**Experiments:** The electrical noise applied to the junction is white Gaussian noise with a bandwidth $F_N = 40\,MHz$. Measurements are performed under an in-plane applied field $H_0$ of 59 Oe in order to compensate the residual stray field produced by the reference layer (synthetic antiferromagnet), and thus equilibrate dwell times in the P and AP states in the absence of applied voltage.

**Data analysis:** In order to determine the resistance of the junction as a function of time, we record the current flowing through the junction with an oscilloscope. As the driven voltage oscillates between two values ($+V_{ac}$ and $-V_{ac}$) and the magnetic tunnel junction switches between two resistance states ($R_{AP}$=640 Ω and $R_P$=390 Ω), the current flowing through the junction can take four values. At high level of electrical noise, the determination of the resistance state of the junction becomes more difficult, as can be seen from the increasing width of error bars in Fig.2a.

**Numerical simulations:** We assume that the free layer of the superparamagnetic tunnel junction can be considered as a single domain magnetization element and follows the Neel-Brown model[25] in which the escape rates of this process are described by Arrhenius equations and can be controlled through the handle of spin transfer torque[26,27]

$$\phi_{AP\to P\ (P\to AP)} = \phi_0 exp\left(-\frac{\Delta E}{k_B T}\left(1 \pm \frac{V(t) + V_N(t)}{V_c}\right)\right)$$

In this study $\phi_0 = 10^9 s^{-1}$ is the effective attempt frequency[27], $\Delta E$ is the energy barrier between the two stable states, $k_B$ is the Boltzmann constant, $T$ is the temperature, $U(t) = V(t) + V_N(t)$ is the applied voltage and $V_c$ is the threshold voltage for deterministic switching [26,27].

For numerical simulations we compute at each time step the probability for the magnetization to switch during the time interval $dt$ knowing the initial state:

$$P_{P \to AP\ (AP \to P)} = 1 - exp(-dt\phi_{P \to AP\ (AP \to P)}(t))$$

A pseudo-random number is then generated to decide whether the switch occurs or not. The parameters of the model: $F_{ac}$, $V_{ac}$ (frequency and amplitude of the driving square voltage) and $F_N$ (bandwidth of the electrical Gaussian noise) have values identical to the experimental protocol. The chosen time step corresponds to the smallest time scale of the experimental noise generator $dt = 1/F_N$. The mean frequency is computed as the mean number of oscillations of the junction per second.

Matching numerical predictions with experimental results for the evolution of the frequency of the magnetic tunnel junction versus the level of noise (Fig. 2(a) left axis) allows us to extract the two free parameters of the model: the ratio $\Delta E/k_B T = 22.5$ and the critical voltage $V_c = 235 mV$.

**Analytical model:** The specificity of external noise is that it introduces a supplementary level of randomness as compared to the internal noise provided by temperature: the escape rates depend on the electrical noise $N(t)$ and are therefore random variables themselves. Therefore, the probability for the magnetization to switch during half a period $T_{ac}/2$ of the drive can only defined as an average over the possible values of N:

$$P_+ = <P_{P \to AP}(+V_{ac})> = <P_{AP \to P}(-V_{ac})>$$
$$P_- = <P_{P \to AP}(-V_{ac})> = <P_{AP \to P}(+V_{ac})>$$

where $P_{P \to AP}(+V_{ac})$ is the probability to switch from P to AP during $T_{ac}/2$ when the excitation voltage is $+V_{ac}$.

Therefore, $P_+$ is the probability to switch from out of phase to in-phase during $T_{ac}/2$ while $P_-$ is the probability to switch from in-phase to out of phase during $T_{ac}/2$.

$$P_\pm = 1 - X_\pm^{\frac{T_{ac}}{2dt}}$$

With $X_\pm = \int_{-\infty}^{+\infty} dN \times \Psi(N) \times \left(1 - exp\left(-dt\phi_0 exp\left(-\frac{\Delta E}{k_B T}\left(1 \pm \frac{V_{ac}+V_N}{V_c}\right)\right)\right)\right)$

and $\Psi(N)$ is a Gaussian distribution over N.

When the level of noise is sub-optimal, synchronization is limited by the junction's ability to phase-lock fast enough when the excitation voltage reverses. Therefore the mean frequency of the junction is $F = F_{ac}(2P_+ - 1)$. On the other hand, when the noise level is supra-optimal, synchronization is limited by the junction's tendency to jump out of phase with the excitation voltage. Therefore $F = F_{ac}(1 + 2P_-)$. On the whole, near the plateau, the mean frequency of the junction is $F = F_{ac}(2P_+ + 2P_- - 1)$. In consequence, $P_+ > 99.5\%$ and $P_- < 0.5\%$ means that the junction is frequency-locked with less than 1% error.

$V_0$ is computed as the minimum voltage drive $V_{ac}$ for which there is an electrical noise level $\sigma_0$ that satisfies $P_+>99.5\%$ and $P_-<0.5\%$. $V_1$ is the minimum voltage drive $V_{ac}$ for which $P_+>99.5\%$ and $P_-<0.5\%$ is satisfied at zero electrical noise.

**Energy consumption predictions:** We consider a perpendicularly magnetized tunnel junction and make the assumption that the free layer is a single magnetization element ($D < 30\ nm$). The thickness of the free layer in the junction is fixed, so that the energy barrier between the two magnetization states scales with the square of the diameter D: $\Delta E = \Delta E_0 \frac{D^2}{D_0^2}$. We also consider that the resistance × area product $RA$ of the junction is constant, so that the resistance of the junction can be written as: $R = \frac{RA}{\pi D^2/4}$. We use numerical parameters from Sato et al.[2]: $\Delta E_0 = 90\ k_B T$, $D_0 = 30\ nm$, $V_c = 0.71\ V$ and $RA = 10\ \Omega.\mu m^2$.

For the energy consumption of a deterministic spin-torque oscillator we considered the same model with a resistance × area product $RA = 3\ \Omega.\mu m^2$ and a diameter of 24nm (which corresponds to the traditionally required energy barrier of $\Delta E = 60\ k_B T$). The power consumption is dominated by the DC current required to induce high frequency oscillations of the magnetization, therefore $P = \frac{V_c^2}{R} \approx 8 \times 10^{-4} W$. Thus for a spin torque oscillator of 10 GHz frequency taking $T_{sync}$ = 10ns to reach synchronization[31] the energy consumption is $E = P \times T_{sync} \approx 8 \times 10^{-12} J$.

1. Khajetoorians, A. A. *et al.* Current-Driven Spin Dynamics of Artificially Constructed Quantum Magnets. *Science* **339,** 55–59 (2013).
2. Sato, H. *et al.* Properties of magnetic tunnel junctions with a MgO/CoFeB/Ta/CoFeB/MgO recording structure down to junction diameter of 11 nm. *Applied Physics Letters* **105,** 062403 (2014).
3. Finocchio, G., Krivorotov, I. N., Cheng, X., Torres, L. & Azzerboni, B. Micromagnetic understanding of stochastic resonance driven by spin-transfer-torque. *Phys. Rev. B* **83,** 134402 (2011).
4. d'Aquino, M., Serpico, C., Bonin, R., Bertotti, G. & Mayergoyz, I. D. Stochastic resonance in noise-induced transitions between self-oscillations and equilibria in spin-valve nanomagnets. *Phys. Rev. B* **84,** 214415 (2011).

5. Cheng, X., Boone, C., Zhu, J. & Krivorotov, I. Nonadiabatic Stochastic Ressonancee of a nanomagnet excited by spin torque. *Phys. Rev. Lett.* (2010).

6. Gammaitoni, L., Hänggi, P., Jung, P. & Marchesoni, F. Stochastic resonance. *Rev. Mod. Phys.* **70,** 223–287 (1998).

7. Locatelli, N., Cros, V. & Grollier, J. Spin-torque building blocks. *Nat Mater* **13,** 11–20 (2014).

8. Locatelli, N. *et al.* Noise-Enhanced Synchronization of Stochastic Magnetic Oscillators. *Phys. Rev. Applied* **2,** 034009 (2014).

9. Pufall, M. *et al.* Physical implementation of coherently-coupled oscillator networks. *IEEE Journal on Exploratory Solid-State Computational Devices and Circuits* **PP,** 1–1 (2015).

10. Chen, A., Hutchby, J., Zhirnov, V. & Bourianoff, G. *Emerging Nanoelectronic Devices*. (John Wiley & Sons, 2014).

11. Yogendra, K., Fan, D. & Roy, K. Coupled Spin Torque Nano Oscillators for Low Power Neural Computation. *IEEE Transactions on Magnetics* **51,** 1–9 (2015).

12. Locatelli, N. *et al.* Spin torque nanodevices for bio-inspired computing. in *2014 14th International Workshop on Cellular Nanoscale Networks and their Applications (CNNA)* 1–2 (2014). doi:10.1109/CNNA.2014.6888659

13. Sharma, A. A., Bain, J. A. & Weldon, J. A. Phase Coupling and Control of Oxide-Based Oscillators for Neuromorphic Computing. *IEEE Journal on Exploratory Solid-State Computational Devices and Circuits* **1,** 58–66 (2015).

14. Corinto, F., Ascoli, A. & Gilli, M. Nonlinear Dynamics of Memristor Oscillators. *IEEE Transactions on Circuits and Systems I: Regular Papers* **58,** 1323–1336 (2011).

15. Aonishi, T. Phase transitions of an oscillator neural network with a standard Hebb learning rule. *Phys. Rev. E* **58,** 4865–4871 (1998).

16. Hoppensteadt, F. C. & Izhikevich, E. M. Oscillatory Neurocomputers with Dynamic Connectivity. *Phys. Rev. Lett.* **82,** 2983–2986 (1999).

17. Fell, J. & Axmacher, N. The role of phase synchronization in memory processes. *Nat Rev Neurosci* **12,** 105–118 (2011).

18. Neiman, A., Schimansky-Geier, L., Moss, F., Shulgin, B. & Collins, J. J. Synchronization of noisy systems by stochastic signals. *Phys. Rev. E* **60,** 284–292 (1999).

19. Shulgin, B., Neiman, A. & Anishchenko, V. Mean Switching Frequency Locking in Stochastic Bistable Systems Driven by a Periodic Force. *Phys. Rev. Lett.* **75,** 4157–4160 (1995).

20. Freund, J. A., Schimansky-Geier, L. & Hänggi, P. Frequency and phase synchronization in stochastic systems. *Chaos: An Interdisciplinary Journal of Nonlinear Science* **13,** 225–238 (2003).

21. Khovanov, I. A. & McClintock, P. V. E. Synchronization of stochastic bistable systems by biperiodic signals. *Phys. Rev. E* **76,** 031122 (2007).

22. Barbay, S., Giacomelli, G., Lepri, S. & Zavatta, A. Experimental study of noise-induced phase synchronization in vertical-cavity lasers. *Phys. Rev. E* **68,** 020101 (2003).

23. Berger, L. Emission of spin waves by a magnetic multilayer traversed by a current. *Phys. Rev. B* **54,** 9353–9358 (1996).

24. Slonczewski, J. C. Excitation of spin waves by an electric current. *Journal of Magnetism and Magnetic Materials* **195,** L261–L268 (1999).

25. Brown, W. F. Thermal Fluctuations of a Single-Domain Particle. *Phys. Rev.* **130,** 1677–1686 (1963).

26. Li, Z. & Zhang, S. Thermally assisted magnetization reversal in the presence of a spin-transfer torque. *Phys. Rev. B* **69,** 134416 (2004).

27. Rippard, W., Heindl, R., Pufall, M., Russek, S. & Kos, A. Thermal relaxation rates of magnetic nanoparticles in the presence of magnetic fields and spin-transfer effects. *Phys. Rev. B* **84,** 064439 (2011).

28. Mizrahi, A. *et al.* Magnetic Stochastic Oscillators: Noise-Induced Synchronization to Underthreshold Excitation and Comprehensive Compact Model. *IEEE Transactions on Magnetics* **51,** 1–4 (2015).

29. Casado-Pascual, J. *et al.* Theory of frequency and phase synchronization in a rocked bistable stochastic system. *Phys. Rev. E* **71,** 011101 (2005).

30. Pikovsky, A., Rosenblum, M. & Kurths, J. *Synchronization: a universal concept in nonlinear sciences*. **12,** (Cambridge university press, 2003).

31. Rippard, W., Pufall, M. & Kos, A. Time required to injection-lock spin torque nanoscale oscillators. *Applied Physics Letters* **103,** 182403 (2013).

32. Livi, P. & Indiveri, G. A current-mode conductance-based silicon neuron for address-event neuromorphic systems. in *IEEE International Symposium on Circuits and Systems, 2009. ISCAS 2009* 2898–2901 (2009). doi:10.1109/ISCAS.2009.5118408

**Acknowledgements:**

The authors acknowledge financial support from the FET-OPEN Bambi project No. 618024 and the ANR MEMOS n° ANR-14-CE26-0021-01. A. M. acknowledges financial support from the Ile-de-France regional government through the DIM nano-K program. R. L acknowledges financial support from the ANR agency (SPINNOVA ANR-11-NANO-0016). N. L. acknowledges financial support from a public grant overseen by the French National Research Agency (ANR) as part of the "Investissements d'Avenir" program (Labex NanoSaclay, reference: ANR-10-LABX-0035).

**Contributions:**

A. F., H. K. and S. Y. designed, optimized, and fabricated the superparamagnetic tunnel junctions. A. M. and R. L conducted the experiments. A. M., N. L, D. Q and J. G. developed the theoretical model and wrote this letter. All authors discussed the results and reviewed the manuscript.

**Additional information**

**Competing interest**: The authors declare no competing financial interests.

**Figure legends**

**Figure 1: Controlling the phase locking of a superparamagnetic tunnel junction through electrical noise: experimental results.** A square periodic voltage of amplitude $V_{ac} = 63$ mV and frequency $F_{ac} = 50$ Hz as well as white Gaussian electrical noise are applied to the junction. (a) Inset: schematic of the superparamagnetic tunnel junction driven by a periodic square voltage and electrical noise. Main: junction's mean frequency as a function of electrical noise amplitude (standard deviation $\sigma_{Noise}$). (b) Times traces of the junction's resistance (top) and applied voltage (bottom) for three different levels of noise with standard deviations: (1) $\sigma_{Noise} = 15$ mV, (2) $\sigma_{Noise} = 30$ mV and (3) $\sigma_{Noise} = 40$ mV

**Figure 2: Modelling the phase locking of superparamagnetic tunnel junctions to an external periodic drive in the presence of electrical noise.** Simulations and analytical calculations are done with the same set of parameters: $V_c$ = 235mV and $\Delta E/k_B T$ = 22.5. (a) A square periodic voltage of frequency $F_{ac} = 50$ Hz and a white Gaussian electrical noise are applied to a magnetic tunnel junction. Three amplitudes are studied: $V_{ac} = 44$ mV (green), $V_{ac} = 50$ mV (blue) and $V_{ac} = 63$ mV (red). Left axis: frequency of the oscillator versus the standard deviation of the noise, both experimental results (circles, squares and triangles) and numerical results (solid lines) are represented. Right axis: analytical values of probabilities $P_+$ and $P_-$ to switch during half a period $T_{ac}/2$ versus noise (dash lines). Vertical dot lines represent the noise levels for which $P_+ = 99.5\%$ and $P_- = 0.5\%$ for a 63 mV amplitude. The horizontal black solid line represents the drive frequency $F_{ac}$. (b-c) Lower noise bound (black) and higher noise

bound (red) of the synchronization plateau versus the drive voltage (b) and versus the drive frequency (c). Both analytical values (dash lines) and experimental results (circles and squares) are presented. In the red zones the oscillator is synchronized with the excitation.

**Figure 3: Energy required to phase-lock a perpendicularly magnetized superparamagnetic tunnel junction: predictions of the analytical model.** Upper inset: schematic of Figure 2b. The circle indicates the lowest drive voltage $V_0$ for which synchronization can be achieved and the corresponding electrical noise level $\sigma_0$. The star indicates the lowest drive voltage $V_1$ for which synchronization can be achieved through thermal noise alone without addition of any electrical noise. Lower inset: schematic view of a perpendicularly magnetized tunnel junction. Main: Calculated minimum energy required to synchronize a perpendicularly magnetized superparamagnetic tunnel junction to a periodic voltage drive in one period, plotted versus the diameter of the junction, for different drive frequencies. Circles represent the case where electrical noise has been added while stars represent the case where only thermal noise is used.